\documentclass[aps,twocolumn,preprintnumbers,final]{revtex4}
\usepackage{lipsum}
\makeatletter
\newcommand*{\balancecolsandclearpage}{%
  \close@column@grid
  \clearpage
  \twocolumngrid
}
\makeatother

\usepackage{graphicx,amsmath,amssymb,mathrsfs,array,longtable,delarray,verbatim,epsfig}
\graphicspath{{E:\My_papers\Multiplexer}}

\begin{document}

\title{Dependence of the 0.7 anomaly on the curvature of the potential barrier in quantum wires}

\author{L. W. Smith$^{1,*,\dagger}$, H. Al-Taie$^{1,2}$, A. A. J. Lesage$^{1}$, F. Sfigakis$^{1}$, P. See$^{3}$, J. P. Griffiths$^{1}$, H. E. Beere$^{1}$, G. A. C. Jones$^{1}$, D. A. Ritchie${^1}$, A. R. Hamilton$^4$, M. J. Kelly$^{1,2}$, and C. G. Smith$^{1}$}

\affiliation{
$^{1}$Cavendish Laboratory, Department of Physics, University of Cambridge, J. J. Thomson Avenue, Cambridge, CB3 0HE, United Kingdom\\
$^{2}$Centre for Advanced Photonics and Electronics, Electrical Engineering Division, Department of Engineering, 9 J. J. Thomson Avenue, University of Cambridge, Cambridge CB3 0FA, United Kingdom\\
$^{3}$National Physical Laboratory, Hampton Road, Teddington, Middlesex TW11 0LW, United Kingdom\\
$^{4}$ School of Physics, University of New South Wales, Sydney NSW 2052, Australia}

\date{\today}
           
\begin{abstract}

Ninety eight one-dimensional channels defined using split gates fabricated on a GaAs/AlGaAs heterostructure are measured during one cooldown at $1.4$ K. The devices are arranged in an array on a single chip, and individually addressed using a multiplexing technique. 
The anomalous conductance feature known as the ``0.7 structure'' is studied using statistical techniques. 
The ensemble of data show that the 0.7 anomaly becomes more pronounced and occurs at lower values as the curvature of the potential barrier in the transport direction decreases.
This corresponds to an increase in the effective length of the device.
The 0.7 anomaly is not strongly influenced by other properties of the conductance related to density.
The curvature of the potential barrier appears to be the primary factor governing the shape of the 0.7 structure at a given $T$ and $B$.

\end{abstract}

\maketitle

\section{Introduction}

Electrons confined in narrow one-dimensional (1D) channels occupy discrete energy levels or subbands, causing the measured conductance to be quantized in units of $G_0=2e^2/h$~\cite{Wharam1988, vanWees1988}. Interactions between the electrons give rise to an anomalous feature near $0.7G_0$, the ``0.7 structure/anomaly''~\cite{Thomas1996}.
Consensus about the origin of the 0.7 anomaly has not yet been reached~\cite{Micolich2011, Micolich2013}, and it has been attributed to various causes. These include spontaneous spin polarization~\cite{Thomas1996, Wang1996}, the Kondo effect~\cite{Cronenwett2002, Meir2002, Rejec2006, Iqbal2013}, Wigner crystallisation~\cite{Matveev2004, Brun2014}, and inelastic scattering combined with a smeared van Hove singularity in the density of states~\cite{Sloggett2008, Bauer2013}.
The difficulty in distinguishing between different scenarios arises in part from understanding how the 0.7 anomaly is governed by specific factors. 
Its behavior as a function of temperature $T$ and magnetic field $B$ is well established: the 0.7 structure becomes more pronounced as $T$ increases (while higher plateaus are smeared by thermal effects), and evolves into a plateau at $0.5G_0$ with increasing $B$, as the 1D subbands are split by the Zeeman energy.

The evolution of the 0.7 structure as a function of carrier density and 1D channel geometry is not well established. As carrier density is reduced, the 0.7 structure has been shown to evolve from $0.7G_0$ to $0.5G_0$~\cite{Thomas1998, Thomas2000, Nuttinck2000}. However, other studies report more complex behavior, with the 0.7 structure rising and/or lowering in conductance as carrier density increases \cite{Pyshkin2000, Reilly2001, Hashimoto2001, Burke2012}. In addition, the only studies of the 0.7 structure as a function of device geometry have shown opposite trends \cite{Iqbal2013, Reilly2001}, or no clear dependence \cite{Koop2007}. The possible interplay between channel length and carrier density further complicates the interpretation of experimental results.

In this paper, we propose a solution to the apparent conflict between reports on the density dependence of the 0.7 structure. Contrary to what is widely assumed in the literature, we believe that it is the field profile, rather than carrier density, that dictates the behavior of the 0.7 structure at constant temperature and in zero magnetic field. More specifically, the primary variable appears to be the longitudinal curvature of the one-dimensional barrier. This hitherto unreported observation is made possible by the use of two recently-developed experimental methods. The first, multiplexing~\cite{Al-Taie2013}, enables a statistical analysis on the characteristics of 98 split gates within the same cooldown (the largest reported study in one-dimensional transport). The second is a data analysis tool that removes the trivial geometry dependence of the conductance, leaving only the contribution of electron-electron interactions. 
Our data are analyzed within a framework given by the van Hove scenario \cite{Bauer2013} for the 0.7 structure, showing good qualitative agreement. Although we cannot at this moment rule out any models of the 0.7 structure/anomaly, we can however formulate a quantitative predictive test which could discriminate between such models.

Studies of transport in split gates have so far relied on measuring individual devices, and reproducing the result in a few others. An entirely different approach is presented here: large numbers of devices are measured in a single cooldown and trends are observed in the combined data. This is achieved using a multiplexing technique~\cite{Al-Taie2013}, which overcomes limitations on the number of devices that can be located on a single chip and measured in cryogenic apparatus. We study an ensemble of data obtained from 98 lithographically-identical split gates, in which differences in potential landscape arise from fluctuations in the background potential. The statistical significance of our data analysis is a departure from methods previously used to study the 0.7 structure. 

The paper is arranged as follows: 
First, details of experimental techniques are given in Sec. II.
Next, Sec. III describes variations in conductance properties between devices, and presents the method of data analysis for removing the geometric dependence of the conductance trace for a non-interacting framework, such that differences due to electron interactions can be directly compared.
The dependence of the 0.7 structure on the barrier curvature is then analyzed in Sec. IV. 
Section V compares our results with the van Hove scenario for the 0.7 structure, and Sec. VI presents an alternative method of analyzing our data (confirming trends observed in Sec. III). This is followed by discussion and conclusions in Secs. VII and  VIII, respectively.

\section{Methods}

\begin{figure}
\includegraphics[width=8cm,height=15cm,keepaspectratio]{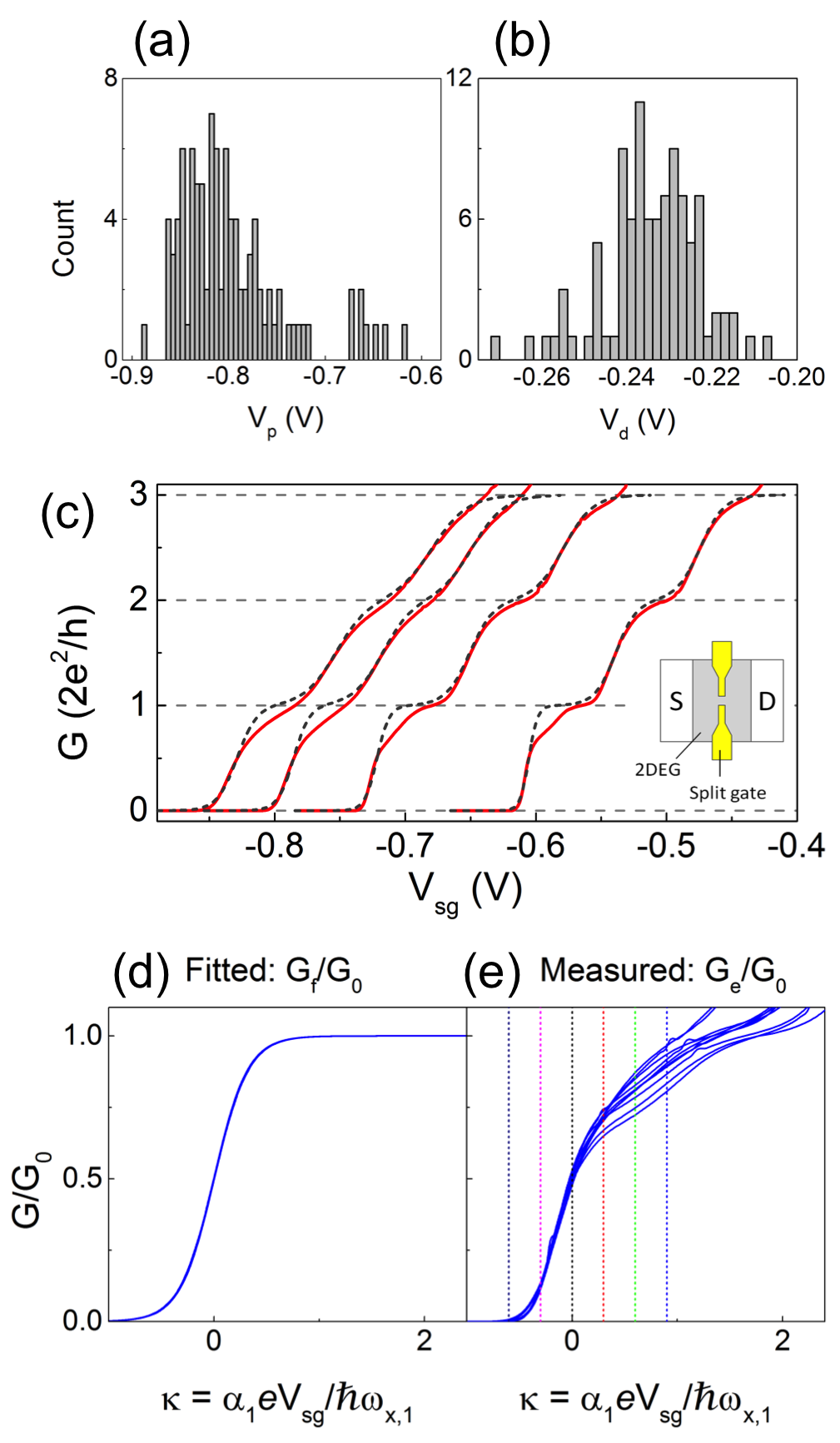}
\caption{\label{Fig1} 
(a), (b) Histograms of pinch-off voltage $V_p$ and 1D definition voltage $V_d$, respectively. Data are shown for 98 split gates. The bin sizes for panels (a) and (b) are 5 mV and 2 mV, respectively.
(c) Conductance $G$ as a function of split-gate voltage $V_{sg}$, for four example devices (solid lines). The dashed lines show fits to the data using a modified saddle-point model, with barrier curvature $\hbar\omega_{x,n}$ as a fitting parameter. 
For the data shown, $\hbar\omega_{x,1}$ decreases from left-to-right (there is no specific relationship between $\hbar\omega_{x,1}$ and pinch-off voltage $V_p$; these traces are chosen to show examples of $\hbar\omega_{x,1}$ at different $V_p$ for clarity).
(d) Fitted conductance $G_f/G_0$ for 98 split gates, collapsed onto a universal curve. The fitting for the first subband is shown. Data are offset to align $G_f/G_0=0.5$ with $V_{sg}=0$ for each trace, and scaled by $\alpha_1 e/\hbar\omega_{x,1}$. 
(e) Corresponding experimentally-measured conductance $G_e/G_0$, for a subset of ten devices. The data are offset and scaled using the same parameters as $G_f$. For $G_e/G_0<0.5$, the traces collapse onto a similar curve. Above $0.5G_0$, variations arise because of differences in the shape of the 0.7 structure. 
Changes in the 1D subband spacing cause a further spread for $G_e/G_0>1$. Small artefacts are present in a few traces due to measurement glitches, and do not affect the analysis.
Six vertical dotted lines show $\kappa$ from $-0.6$ to $0.9$ in steps of $0.3$. }
\end{figure}

An array of 256 split gates is fabricated on a GaAs/AlGaAs modulation-doped heterostructure, in which the two-dimensional electron gas (2DEG) forms 90 nm below the surface of the wafer. The carrier density and mobility are measured to be $1.7\times10^{11}$ cm$^{-2}$ and $0.94\times10^6$ cm$^2$V$^{-1}$s$^{-1}$, respectively. 
Each split gate defines a 1D channel in the underlying 2DEG~\cite{Thornton1986}, and is defined by electron-beam lithography to be $0.4$ $\mu$m long and $0.4$ $\mu$m wide. The inset of Fig.~\ref{Fig1}(c) shows a schematic of a split-gate device. Other surface gates are defined by optical lithography, and all gates are metallized by thermally evaporating Ti/Au. 
The differential conductance is measured using two-terminal, constant voltage methods, with an ac excitation voltage of 100 $\mu$V at 77 Hz.
The measurements were performed at $T=1.4$ K, in order for the 0.7 structure to be well developed.

Each split gate is measured individually (details of this technique are given in Refs.~\cite{Al-Taie2013, Al-Taie2013B}).
Data showing strong evidence of disorder for low $G$ (less than $3G_0$) were discarded, where disorder effects manifest in deviations in the quantization of conductance (such as an absence of conductance plateaus or suppression below the expected value), or the existence of resonant features. Data from 98 split gates are retained and corrected for series resistance using the measured resistance at $V_{sg} = 0$ as a systematic method of analyzing the data.

\section{Variations of conductance properties}

Despite the identical lithographic geometry, differences exist in the 1D conductance properties (an initial study of correlations between these properties is reported in Ref.~\cite{Smith2014}).
Figures \ref{Fig1}(a) and \ref{Fig1}(b) show histograms of the pinch-off voltage of the 1D channel $V_p$ and the voltage at which the 1D channel is first defined $V_d$~\cite{note1}, respectively. Data from all 98 split gates are presented.
The spread of these parameters arises from changes in the potential landscape predominantly due to the fluctuating background potential from ionized dopants and impurities (it is likely that for the 98 split gates considered, the impurities are not close to the 1D channel since the transmission through the first three 1D subbands is not affected).

Figure~\ref{Fig1}(c) shows conductance $G$ as a function of split-gate voltage $V_{sg}$ for four devices (solid lines). Fits to the data are shown by dashed curves.
The fitted conductance is calculated using the Landauer-B\"{u}ttiker formalism for non-interacting electrons traversing a modified saddle-point potential~\cite{Buttiker1990}. This has transmission probability $T_n = [1+\exp(-2\pi(E-E_n)/\hbar\omega_{x,n})]^{-1}$, where $E_n$ is the energy of the subband bottom at the center of the 1D channel, $\hbar\omega_{x,n}$ is the curvature of the barrier in the transport direction, and $n$ is the subband index. Subband-dependent values of $\hbar\omega_{x,n}$ achieve a better overall fit, which reflects a modification of the barrier curvature due to changes in electron density.
We refer to the experimentally-measured conductance as $G_e$, and fitted conductance as $G_f$.
The fitting procedure is described in the appendix.

For split gates, the conductance below $0.5G_0$ is independent of temperature up to intermediate $T$ (typically $\approx1.5$ K for devices similar to those used here~\cite{Thomas1996}).
Therefore, the fitting is performed with $T=0$ for $n=1$ ($T=1.4$ K for $n=2$ and $3$, since here $G$ is sensitive to $T$).  
We have repeated the analysis with $T=1.4$ K for $n=1$, which shows the same trends. These data are shown in the Supplemental Material~\cite{SuppMat}. To our knowledge, the anomalous behavior of the first transition width has not been studied in detail in experiment. However, the van Hove scenario for the 0.7 structure gives detailed predictions for its behavior (discussed in Sec.~\ref{vanHove}). In order to test predictions of the van Hove scenario regarding the 0.7 structure our data are analyzed within its framework.

Parameter $\hbar\omega_{x,1}$ governs the width in gate voltage of the transition in $G$ from zero to $G_0$. 
Variations in transition widths for different devices are removed by scaling the gate voltage by $\alpha_1 e/\hbar\omega_{x,1}$,
where $e$ is the electronic charge, and $\alpha_1$ is an average lever arm obtained from dc bias spectroscopy measurements (the method for obtaining $\alpha$ is described in the Appendix).
Prior to scaling the traces are offset horizontally to align $G_f/G_0=0.5$ with $V_{sg}=0$.
Figure~\ref{Fig1}(d) shows the fitted conductance curves $G_f$ as a function of $\kappa=\alpha_1 eV_{sg}/\hbar\omega_{x,1}$ for the first subband, which collapses the data set of 98 traces onto a single line (akin to Fig. S14 of \cite{Bauer2013}).

The same procedure is applied to the experimental data $G_e/G_0$, shown in Fig.~\ref{Fig1}(e). For clarity, only ten traces are shown.
Below $G_e=0.5G_0$, the data collapse onto a similar curve. 
However, variations arise above $G_e=0.5G_0$, due to differences in the shape of the 0.7 structure.
Such variations do not occur in $G_f$ [Fig.~\ref{Fig1}(d)], in which electron interactions are not accounted for.

\section{Analysis of the 0.7 structure}

\begin{figure}
\includegraphics[width=16cm,height=11cm,keepaspectratio]{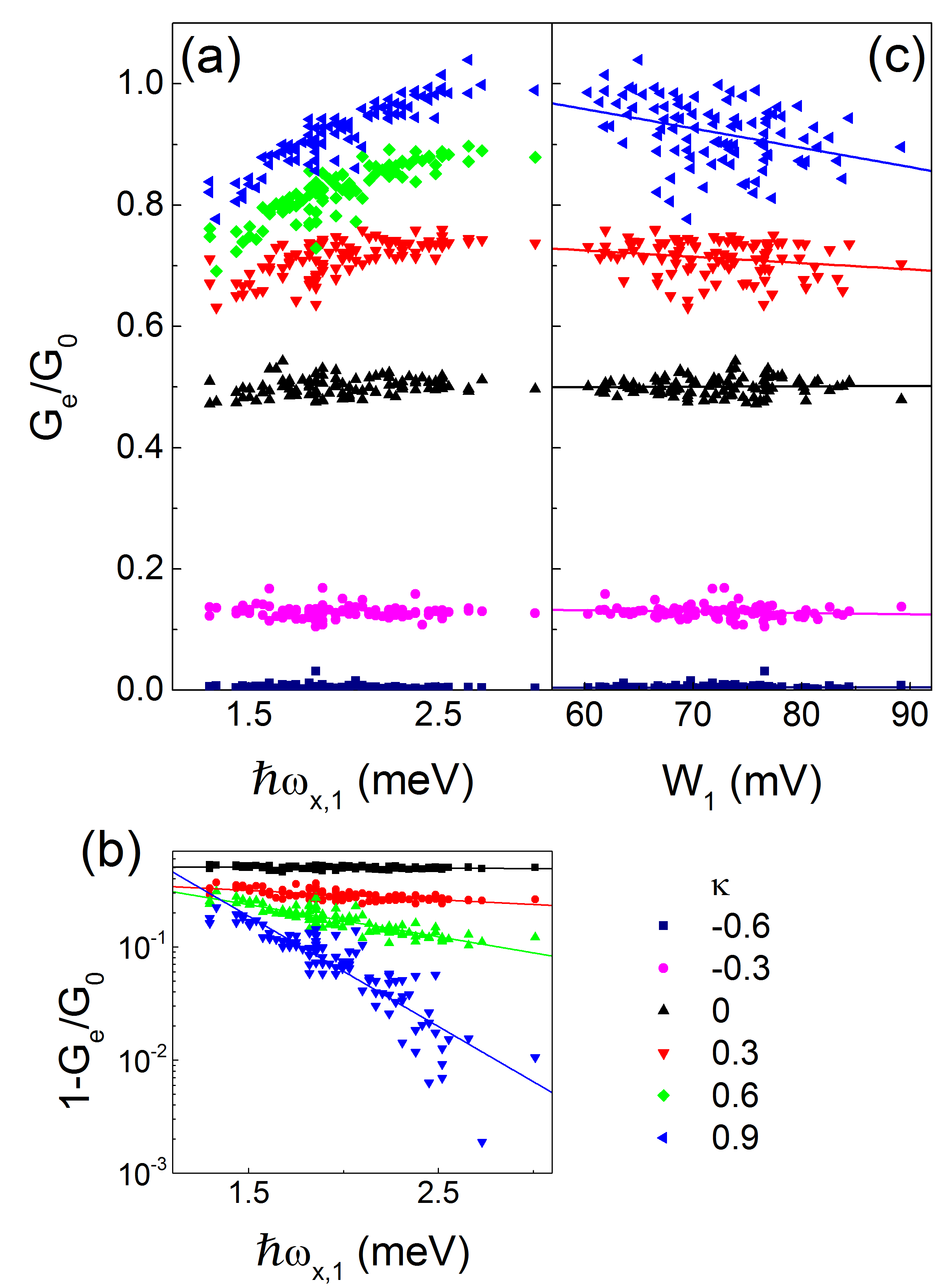}
\caption{\label{Fig2}
Evolution of the 0.7 structure as a function of barrier curvature and plateau width.
(a) $G_e/G_0$ for 98 split gates as a function of $\hbar\omega_{x,1}$, for $\kappa$ from $-0.6$ to $0.9$ in steps of $\Delta \kappa=0.3$ [these $\kappa$ values are shown by vertical dotted lines in Fig.~\ref{Fig1}(e)].
The difference of $G_e/G_0$ from unity is plotted on a linear-log scale in panel (b), for $\kappa = 0$ and above. The solid lines show linear fits to the data, for which the relationship between $G_e$ and $\hbar\omega_{x,1}$ is exponential.
(c) Conductance $G_e/G_0$ as a function of width of the first conductance plateau $W_1$, where $W_1=\Delta V_{sg}$ between $G=0.5$ and $1.5G_0$. 
Data are shown for five fixed values of $\kappa$ ($\kappa = 0.6$ is omitted for clarity). Linear fits to the data are plotted for each $\kappa$. }
\end{figure}

The experimental conductance can now be directly compared at fixed $\kappa$.
Figure~\ref{Fig2}(a) shows $G_e/G_0$ as a function of $\hbar\omega_{x,1}$, at six values of $\kappa$. From bottom-to-top, these correspond to $\kappa=-0.6$ to $0.9$ ($\Delta\kappa=0.3$), illustrated by vertical dotted lines in Fig.~\ref{Fig1}(e), left-to-right.
For $G_e/G_0$ below $\approx 0.6$, $G_e$ does not change with $\hbar\omega_{x,1}$.

For positive $\kappa$, $G_e/G_0$ shows a strong non-linear reduction with decreasing $\hbar\omega_{x,1}$.
This illustrates how the 0.7 anomaly occurs at lower conductance values as the barrier becomes wider, and the effective device length increases.
To investigate the form of this dependence, Fig.~\ref{Fig2}(b) shows a log-linear plot of $1-G_e/G_0$ as a function of $\hbar\omega_{x,1}$ (for $\kappa=0$ to $0.9$). There is an approximately linear trend (solid lines indicate a linear least-squares fit). 
The range of $\hbar\omega_{x,1}$ is small, so that it is difficult to unambiguously identify the specific functional form of the trend. 

Figure~\ref{Fig2}(c) shows $G_e/G_0$ as a function of the width of the first conductance plateau $W_1$. 
For $\kappa \leq 0$, $G_e/G_0$ does not change with $W_1$.
However, for positive $\kappa$ the vertical spread becomes much larger ($\kappa=0.6$ is not plotted for clarity), and there is a weak downward trend as $W_1$ increases. The solid lines in Fig.~\ref{Fig2}(c) show linear least-squares fits to the data. 

This relationship may reflect an additional conductance due to the population of the second 1D subband, since we define $W_1$ by $\Delta V_{sg}$ required for the second subband to fill. The trend is less significant than that observed in Fig.~\ref{Fig2}(a), in which the data are more tightly distributed. Therefore, the trend in Fig.~\ref{Fig2}(a) cannot be attributed to the presence of the second subband.
The decrease in $G_e$ with larger $W_1$ [Fig~\ref{Fig2}(c)] may also be related to the interaction strength, since longer plateaus suggest a stronger transverse confinement. The 0.7 structure is expected to occur at lower conductance values for stronger interactions~\cite{Bauer2013}.

We do not observe any correlations between $G_e/G_0$ and other properties of the conductance trace (including the pinch-off voltage, and the voltage or conductance at which the 1D channel is first defined). These data are shown in Supplementa Material. 
The absence of correlations with other properties of the conductance trace point towards the curvature of the barrier being the primary factor governing the value of the 0.7 structure at a given $T$ and $B$.

Previous studies of the dependence of the 0.7 structure on 2DEG density~\cite{Thomas1998, Thomas2000} have shown conflicting trends. Data from nine studies are summarized in Ref.~\cite{Burke2012}, showing that the 0.7 structure is highly sensitive to the 1D confining potential. It is likely that  the specific potential landscape varies between these devices, due to differences in structure, material, and geometry. It is also likely that the potential changes differently as a function of density for each case, explaining the opposite trends reported.
Our study suggests a specific relationship between the 0.7 structure and the 1D confining potential, and is consistent with data in Ref.~\cite{Reilly2001}, where the 0.7 structure was studied in three devices of different length.

\section{\label{vanHove}Comparison with the inelastic scattering scenario for the 0.7 structure}

We now directly compare our data with the theoretical predictions from Bauer \emph{et al.}~\cite{Bauer2013},
who have argued that a smeared van Hove singularity in the density of states just above the barrier top enhances interaction effects, and thereby causes the 0.7 anomaly \cite{Sloggett2008}.
This is the only model for the origin of this structure for which conductance $g$ has been calculated as a function of $\Omega_x$ (we adopt their notation when referring to this model). Here, $\Omega_x$ is the `bare' barrier curvature, unmodified by electron interactions. 

\begin{figure}
\includegraphics[width=8.5cm,height=8cm,keepaspectratio]{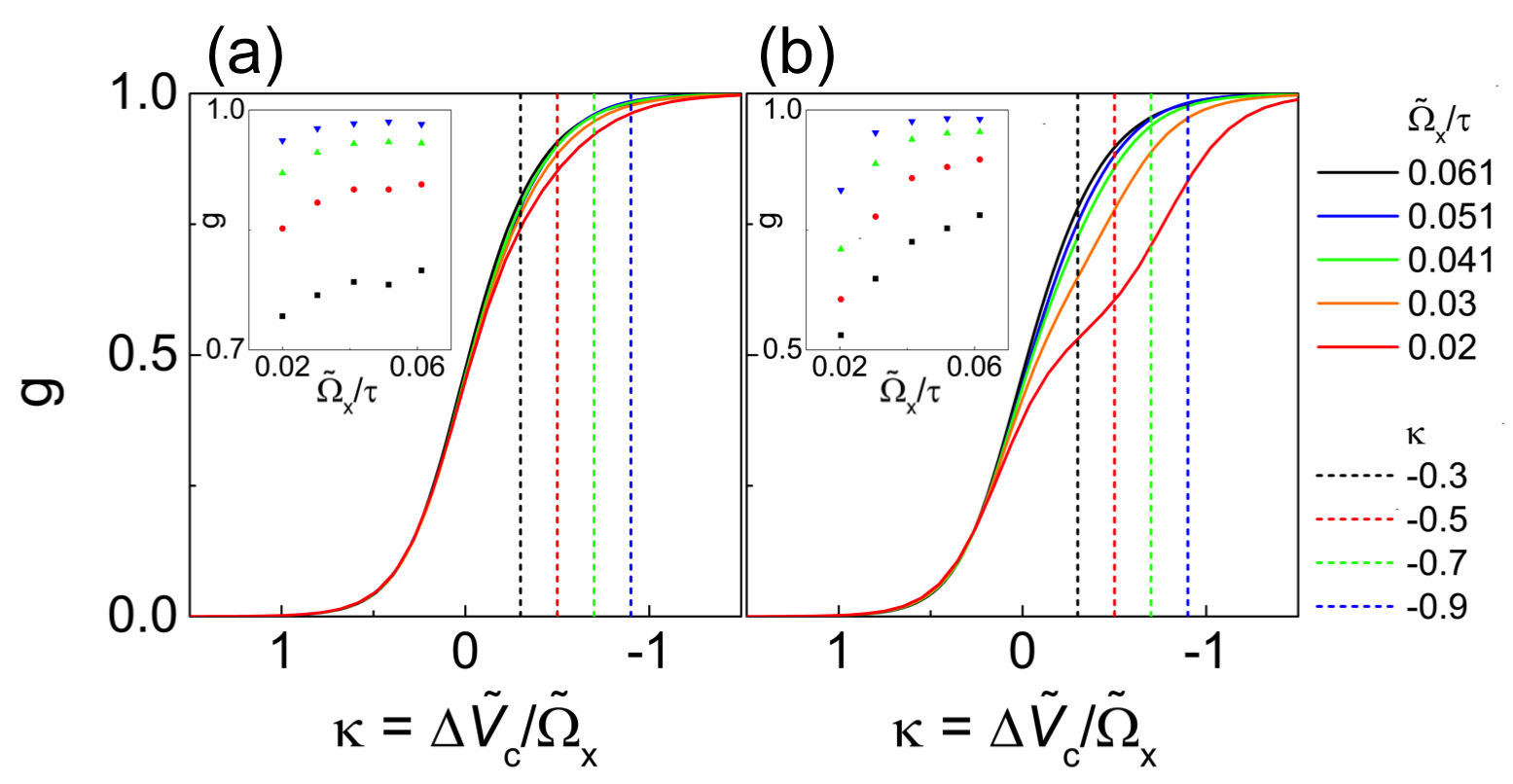}
\caption{\label{Fig3} 
(a) Predicted behavior of conductance $g$ as a function of effective barrier curvature $\tilde{\Omega}_x$, similar to Fig. S14(b) within the supplementary material of Ref.~\cite{Bauer2013}.
Data for five difference values of $\tilde{\Omega}_x/\tau$ are shown, where $\tau$ is the hopping matrix element in the tight-binding model used by Bauer \emph{et al.}~\cite{Bauer2013} (for the case of a half-filled band studied there, $\tau = \epsilon_F/2$, where $\epsilon_F$ is the Fermi energy).
Values of $\tilde{\Omega}_x$ are chosen to collapse the curves onto a single trace for low $g$.
The inset shows the evolution of $g$ as a function of $\tilde{\Omega}_x$ at four different fixed values of $\kappa$. 
From bottom-to-top, these correspond to vertical dashed lines (left-to-right) in the main figure. The data show a reduction in $g$ as $\tilde{\Omega}_x$ is decreased for constant $\kappa$. For data in panel (a) $B=0$ T.
(b) Corresponding dependence of $g$ as a function of $\tilde{\Omega}_x$ for a fixed magnetic field ($B \simeq B_*$ for $\tilde{\Omega}_x = 0.02$).
Calculations are performed for the same five values of the bare barrier curvature $\tilde{\Omega_x}$ as in (a).
The inset shows the evolution of $g$ as a function of $\tilde{\Omega}_x$ for the same four fixed values of $\kappa$.
The reduction in $g$ as $\tilde{\Omega}_x$ is decreased is enhanced compared to the $B=0$ case.
}
\end{figure}

\subsection{Predicted dependence of the 0.7 structure on barrier curvature}

Figure~\ref{Fig3}(a) shows $g$ as a function of $\kappa = \Delta \tilde{V_c}/\tilde{\Omega}_x$, where $\Delta \tilde{V}_c$ is the energy scale 
(offset such that traces align at low $g$), and $\tilde{\Omega}_x$ is an effective barrier curvature which accounts for how the barrier shape is modified by the presence of electrons~\cite{NoteOmega}.
Data for five different values of $\Omega_x$ are calculated using the method for numerical calculations used in the main text of Ref.~\cite{Bauer2013}. They present a more accurate version of Fig. S14(b) within their supplementary information, with slightly different choices for the parameters $U$ and $\Omega_x$. 
In Fig.~\ref{Fig3}(a), $\Delta \tilde{V}_c$ is divided by $\tilde{\Omega}_x$ to collapse the traces onto a single curve near pinch off. 

When $\Omega_x$ is decreased for given $\kappa$, $g$ decreases. The inset shows $g$ as a function of $\tilde{\Omega}_x$ for four values of $\kappa$ (bottom-to-top corresponds to dashed vertical lines left-to-right in the main figure). 
The conductance $g$ decreases in a non-linear fashion with decreasing $\Omega_x$, qualitatively consistent with our measured data in Fig.~\ref{Fig2}(a).

This trend arises because the effective strength of interactions in this model goes as $1/\sqrt{\Omega_x}$, leading to a reduction in $g$ for lower $\Omega_x$~\cite{Bauer2013}. 
These calculations are performed at $T = 0$, therefore the predicted trend is weak. This is also why the model shows a only `skewing' of the conductance trace as $\Omega_x$ reduces, rather than the appearance of the plateau-like shoulder seen in experimental data. 
For this reason, the largest $\Delta g$ occurs for $\kappa$ nearest to zero for the model, whereas in experimental data the largest $\Delta G_e$ occurs at the right-most value of $\kappa$. 

\subsection{Finite magnetic field}

Calculations at non-zero $T$ are currently unavailable, however, at finite $T$ the effects of interactions are enhanced due to increased inelastic scattering, and the model is expected to yield a greater reduction in $g$ with decreasing $\Omega_x$. 
The expected trend can be illustrated (at $T=0$) by considering the case of non-zero $B$, where interaction effects are likewise more pronounced. 
We therefore consider the predicted behavior of the 0.7 anomaly in the van Hove ridge scenario~\cite{Bauer2013} when a constant magnetic field is applied.
Calculations are performed for five difference values of the `bare' barrier curvature $\Omega_x$. 
Figure \ref{Fig3}(b) shows conductance $g$ against $\kappa = \Delta \tilde{V_c}/\tilde{\Omega}_x$, where $\tilde{V}_c$ is the barrier height with respect to the chemical potential ($\tilde{V}_c$ is offset such that traces align at low $g$), and $\tilde{\Omega}_x$ is an effective barrier curvature which accounts for how the barrier shape is modified by the presence of electrons.
The magnetic field $B$ is fixed at a value close to $B_*$ when $\tilde{\Omega}_x = 0.02$ (see Eq. 2 of Ref.~\cite{Bauer2013} for a definition of $B_*$).

For a given $\kappa$, $g$ reduces when $\Omega_x$ is decreased. This is illustrated in the inset, which shows $g$ as a function of $\tilde{\Omega}_x$ for four values of $\kappa$ (bottom-to-top correspond to dashed vertical lines left-to-right in the main figure). 
The fact that the shape of the conductance curve is modified towards lower $g$ with decreasing $\Omega_x$ illustrates that the interaction-induced reduction of conductance becomes more pronounced for lower $\Omega_x$.
This is the same trend as shown in Fig. \ref{Fig3}(a), where $B=0$. The trend is much more pronounced in Fig.~\ref{Fig3}(b), since the interaction-induced reduction of $g$ is strongly enhanced for nonzero $B$.

Within the van Hove ridge scenario, a similar effect is expected for nonzero $T$ at $B=0$ (though no functional renormalization group calculations are available for this case). 
The reduction in $g$ with decreasing $\Omega_x$ is likewise expected to be much more pronounced at $T\neq 0$ compared to $T=0$ [making the size of the trend in Fig. \ref{Fig3}(a) more comparable with experimental data from Fig. \ref{Fig2}(a)].
The reason for this similarity is that both nonzero $B$ and nonzero $T$ amplify the interaction-induced reduction of $g$ (albeit for somewhat different reasons: 
A nonzero magnetic field reduces the conductance due to an interaction-enhanced asymmetry between the effective barrier heights for spin up and spin down electrons, whereas a nonzero temperature reduces the conductance due to enhanced inelastic scattering).

\subsection{Predictions of the behavior of the conductance transition from zero to $0.5G_0$}

Our results are used to test predictions of the inelastic scattering plus van Hove model for the 0.7 structure. This model gives detailed predictions of the behavior of the transition from $G=0$ to $0.5G_0$. Within this scenario, the conductance below $0.5G_0$ is only weakly dependent on interaction strength $U$ in the regime where $B$, $T$, and $V_{sd}$ (and their effective crossover scales; $\tilde{B}_*$, $\tilde{T}_*$, and  $\tilde{V}_{sd*}$, respectively), are small with respect to $\Omega_x$.
This is illustrated in Fig. 1(k) of Ref.~\cite{Bauer2013}, which shows how $g$ evolves with increasing $U$ for  $B$, $T$, and $V_{sd}$ equal to zero. Changing $U$ has only a small effect on the traces for $g<0.5$, such that high $U$ is close to non-interacting scenario for $g<0.5$. 

Related to this, the model also predicts also predicts a weak temperature dependence for $g<0.5$, consistent with experimental data. Figure 2(d) of Ref.~\cite{Bauer2013} shows the predicted behavior of the conductance for various $\tilde{T}/\tilde{T}_*$, where $\tilde{T}_*$ is an effective crossover scale. For $g$ below 0.5, the calculated conductance is almost independent of $\tilde{T}_*$ for $\tilde{T}<\tilde{T}_*$ (for these data $\tilde{T}_*/\Omega_x = 0.2$).

Therefore, for the inelastic scattering plus van Hove scenario the curve shape for $g<0.5$ and $\tilde{T}<\tilde{T}_*$ is governed essentially by $\Omega_x$, such that it is reasonable to use a non-interacting model to analyze our data within the framework of this model, as long as $kT$ is significantly smaller than $\hbar\omega_{x,1}$. Our analysis estimates that $\hbar\omega_{x,1}$ varies from 1.3 to 3 meV for the first subband, thus the maximum ratio of $kT/\hbar\omega_{x,1}$ is $<0.1$.

\section{Alternative analysis of the 0.7 structure}

\begin{figure}
\includegraphics[width=8.5cm,height=9cm,keepaspectratio]{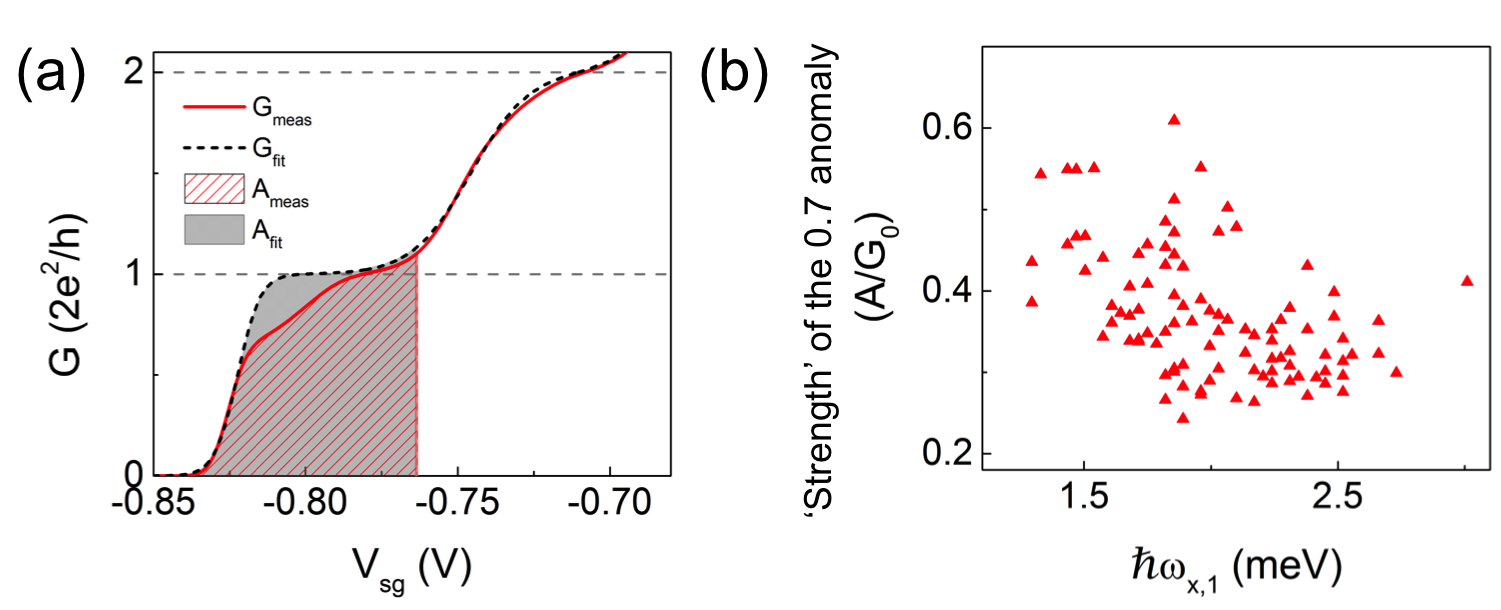}
\caption{\label{Fig4} 
(a) Conductance $G_{meas}$ as a function of $V_{sg}$ for an example split gate (solid line). The area beneath the first plateau is highlighted by the hatched area $A_{meas}$, defined by limits $V_p$ and $V_{sg}$ at $G_{meas}=1.1 G_0$.
The dotted line shows a fit to the data, using a modified saddle-point model. The corresponding area under this curve is shown by the solid pattern ($A_{fit}$). 
(b) Scatter plot of the `strength' of the 0.7 anomaly against barrier curvature $\hbar\omega_{x,1}$, where the strength is estimated using $A = A_{meas} - A_{fit}$. }
\end{figure}

Here we present an alternative method of analyzing the 0.7 anomaly, in terms of its area. The 0.7 structure becomes more pronounced as $\hbar\omega_{x,1}$ reduces, verifying the trend in Fig.~\ref{Fig2}(a).
Figure \ref{Fig4}(a) shows the conductance $G$ as a function of $V_{sg}$ for an example split gate (solid line). The dashed line shows the fitted conductance using the Landauer-B\"{u}ttiker formalism. The difference between the fitted and experimental conductance quantifies how pronounced the 0.7 anomaly appears. 

We define the `strength' of the 0.7 anomaly as $A_{fit}-A_{meas}$, where $A_{meas}$ ($A_{fit}$) is the area beneath the first plateau in measured (fitted) data, respectively. The left and right limits of both areas are given by pinch-off voltage $V_p$ and $V_{sg}$ at $G=1.1G_0$ from the experimental data, respectively. This is illustrated in Fig. \ref{Fig4}(a), in which $A_{meas}$ ($A_{fit}$) is shown by the hatched (solid) pattern.

Figure \ref{Fig4}(b) shows a scatter plot of the strength of the 0.7 anomaly against $\hbar\omega_{x,1}$. There is a strong negative trend: Low $\hbar\omega_{x,1}$ is associated with a stronger 0.7 anomaly. This is consistent with the trend in Fig. \ref{Fig2}(a), where $G$ is reduced as $\hbar\omega_{x,1}$ decreases, since as the 0.7 structure becomes more pronounced (for instance with increasing $T$), the conductance value of the onset of the 0.7 anomaly reduces~\cite{Thomas1996}. This shows that the trend in Fig. \ref{Fig2}(a) is not a result of the scaling of the voltage axis, since no scaling is performed for Fig. \ref{Fig4}(b).

\section{Discussion}

The results presented here do not identify a specific origin of the 0.7 structure, however, they clearly show conditions for which it is strongest. Reproducing the behavior reported here provides a further test of the various theories regarding the 0.7 structure (e.g. Refs.~\cite{Wang1996, Cronenwett2002, Meir2002, Rejec2006, Iqbal2013, Matveev2004, Brun2014, Bauer2013, Sloggett2008, Lunde2009, Aryanpour2009}), in addition to the dependence on magnetic field, temperature and dc bias.

A reduction in $G$ for lower $\hbar\omega_x$ is also predicted by the spontaneous spin polarization~\cite{Jaksch2006} and Kondo scenarios~\cite{Hirose2003}. In the latter case, the relationship between $G$ and $\omega_x$ could be expected to be a power law \cite{NoteKondo}.
However, tests can be devised which may be able to distinguish between the different theories.
For example, the $G_e/G_0$ dependence on $\hbar\omega_{x,1}$ should be investigated as a function of $T$, $B$, and dc source-drain bias $V_{sd}$, which all affect the conductance of the 0.7 structure.
These measurements can be used to extract the low energy scales $\tilde{B}_*$, $\tilde{T}_*$, and $\tilde{V}_{sd*}$ identified in Ref.~\cite{Bauer2013}.
Verifying the predicted dependence of $\tilde{B}_*$ on $\Omega_x$ given in Eq. 35(a) and Fig. S13 of Ref.~\cite{Bauer2013} would be strong evidence that this is the correct model for the 0.7 anomaly.\\

An initial study of correlations between conductance properties of split gates is presented in Ref.~\cite{Smith2014}, using the same data as the current paper. In Ref.~\cite{Smith2014}, we investigated the relationship between pinch-off voltage, the voltage and conductance of the definition point, the width of the first plateau, and barrier curvatures $\hbar\omega_{x,n}$, of subbands $n=1$, 2 and 3. The 0.7 anomaly was also studied. A conductance value ($G_{0.7}$) associated with the 0.7 anomaly was estimated for 36 devices, where $G_{0.7}$ was defined by a local minimum in $dG/dV_{sg}$. No correlations where observed between $G_{0.7}$ and other properties of the 1D conductance. 

The analysis of the 0.7 anomaly in the current study is superior for two reasons. First, this method allows us to investigate the 0.7 anomaly in many more devices. The 0.7 structure is present in all 98 split gates measured but gives rise to a minimum in $dG/dV_{sg}$ for only approximately one third of the data (36 devices); $G_{0.7}$ could be estimated only when the 0.7 structure is very pronounced. Second, in Ref.~\cite{Smith2014} a single conductance value associated with the 0.7 anomaly is found for each device. The approach used in the current study allows the conductances over the entire transition region from zero to $G_0$ to be compared. This provides much more information since the full shape of the suppression of conductance below $G_0$, which forms the 0.7 anomaly, can be studied.

\section{Conclusion}

In summary, we have used a multiplexing technique to systematically study the effect of device geometry on the 0.7 anomaly. Statistical methods were used to analyze a data set from 98 split gates, individually measured during a single cooldown. Trends are identified in the ensemble of data, rather than in the behavior of a single device.
Using an approach to data analysis informed by a recently proposed model for the 0.7 structure, it appears that the 0.7 structure becomes more pronounced and occurs at a lower conductance value for smaller barrier curvature $\hbar\omega_x$. This corresponds to an increase in the effective length of the device, for which the strength of electron interactions increases. Our data suggest that the barrier shape--rather than density--is the primary factor governing the conductance of the 0.7 structure.

\section*{ACKNOWLEDGEMENTS}

This work was supported by the Engineering and Physical Sciences Research Council Grant No. EP/I014268/1. The dataset for this paper is available at www.repository.cam.ac.uk/handle/1810/247687. We thank J. von Delft, S. Ludwig, F. Bauer and J. Heyder for many useful discussions and for providing the data for Fig.~\ref{Fig3}. We thank K. J. Thomas, K.-F. Berggren, T.-M. Chen, E. T. Owen, and M. Pepper for helpful discussions, and R. D. Hall for e-beam exposure.

\section*{APPENDIX}

The fitting routine used to estimate $G_{f}$ is described below. A description can also be found in Ref.~\cite{Smith2014}.

The Landauer-B\"{u}ttiker formalism is used to calculate the conductance as a function of energy $E$ for each subband individually. 
We use a modified saddle-point potential~\cite{Buttiker1990} with transmission probability $T_n = [1+\exp(-2\pi(E-E_n)/\hbar\omega_{x,n})]^{-1}$, where $E_n$ is the energy of the subband bottom at the centre of the 1D channel, and $n$ is the subband index.

The conductance for each subband is calculated independently using 
\begin{equation}
G_n = G_0 \int\;T_n\; \left(-\frac{\partial f}{\partial E}\right)\;dE\ ,
\end{equation}
where $f$ is the Fermi-Dirac distribution $f=(1+e^{(E-\mu)/k_BT})^{-1}$, and $\mu$ is the chemical potential.
In our reference frame each subband edge is initially at $\mu = 0$, such that $E_n = 0$. The integration is performed between  $\pm50k_BT$.

After calculating $G_n$ the energy scale is divided by lever arm $\alpha_n$ ($\Delta E=\alpha e \Delta V_{sg}$, where $e$ is the electronic charge), to convert to a voltage scale. An average value of $\alpha_n$ is used, discussed below.
After scaling, $G_{n}$ for each subband is offset such that $G_{n}=0.5G_0$ aligns with the midpoints of equivalent risers between plateaus in measured conductance ($G_e$). Parameters $\hbar\omega_{x,n}$ are then optimized to fit $G_{n}$ to $G_e$. 
For $n=1$, the fit is performed between 0 and $0.5G_0$ (to avoid the 0.7 structure). For $n=2$ (3), the fitting is performed between $1.01G_0$ and $2G_0$ ($2.01G_0$ and $3G_0$).

As discussed in the main article, the conductance below $0.5G_0$ is independent of temperature up to intermediate $T$. The fitting is therefore performed with $T=0$ for $n=1$ ($T=1.4$ K for $n=2$ and $3$, where $G$ is sensitive to $T$). 
The sum of $G_{n}$ for $n=1$, 2, and 3 gives the final fitted conductance, as shown in Fig.~\ref{Fig1}(c) [the fitted condutance $G_f$ in Fig. \ref{Fig1}(d) is shown for the first subband ($n=1$), since the focus of this article is the 0.7 structure].

DC bias spectroscopy measurements are used to estimate the lever arm $\alpha=\partial V_{sd}/\partial V_{sg}$~\cite{Srinivasan2013}, and the spacing between the 1D subbands $\Delta E_{n,n+1}$. We measured 24 devices, and for the first subband obtain an average $\alpha_1=63\times 10^{-3}$ and $\Delta E_{1,2}=2.8$ meV, with standard deviations of $\alpha_1=8.6\times 10^{-3}$ and $0.3$ meV, respectively~\cite{NB2}.
Because of the time required to perform detailed dc measurements, we measured $G$ for dc-bias voltages from $V_{dc}=0$ to $-2.5$ mV, with coarse $\Delta V_{dc}$ intervals of $-0.5$ mV.
The 1D subband spacing $\Delta E_{1,2}$ is estimated by extrapolating lines linking peaks in the transconductance $dG/dV_{sg}$, following the method used in Ref.~\cite{Patel1991} (see Figs. 2 and 3 of that article).
In order to correct the $V_{dc}$ scale for series resistance we use $V_{dc_{corr}} = V_{dc} \times R_{1D} / (R_{1D} + R_s)$, where $R_{1D}$ is the resistance across the split gate, and $R_s$ is the series resistance.

$*$ Corresponding author: lsmith34@wisc.edu.

$\dagger$ Present address: Department of Physics, University of Wisconsin-Madison, Madison, WI 53711, USA.

\balancecolsandclearpage

\section{Supplemental Material}

Additional material is provided here, supplementary to the main article.
In Sec. A we consider the dependence of the 0.7 anomaly on electrical properties of 1D conductance trace, namely the pinch-off voltage ($V_p$), the voltage at which the 1D channel is defined ($V_d$), and the conductance at which the 1D channel is defined ($G_d$).
In the main article, the fitted conductance for the first subband is calculated using $T=0$ K: In Sec. B, we show that repeating the analysis using $T=1.4$ K (the $T$ at which measurements are carried out) yields very similar results.
Figures presented in this supplementary material (within the main article) are referred to using an ``S'' (``A'') to precede the figure number; e.g. Fig. S1(a) [Fig. A1(a)].

\subsection{Conductance of the 0.7 anomaly as a function of electrical properties of the 1D system}

\begin{figure}[b]
\includegraphics[width=8.4cm,height=10cm,keepaspectratio]{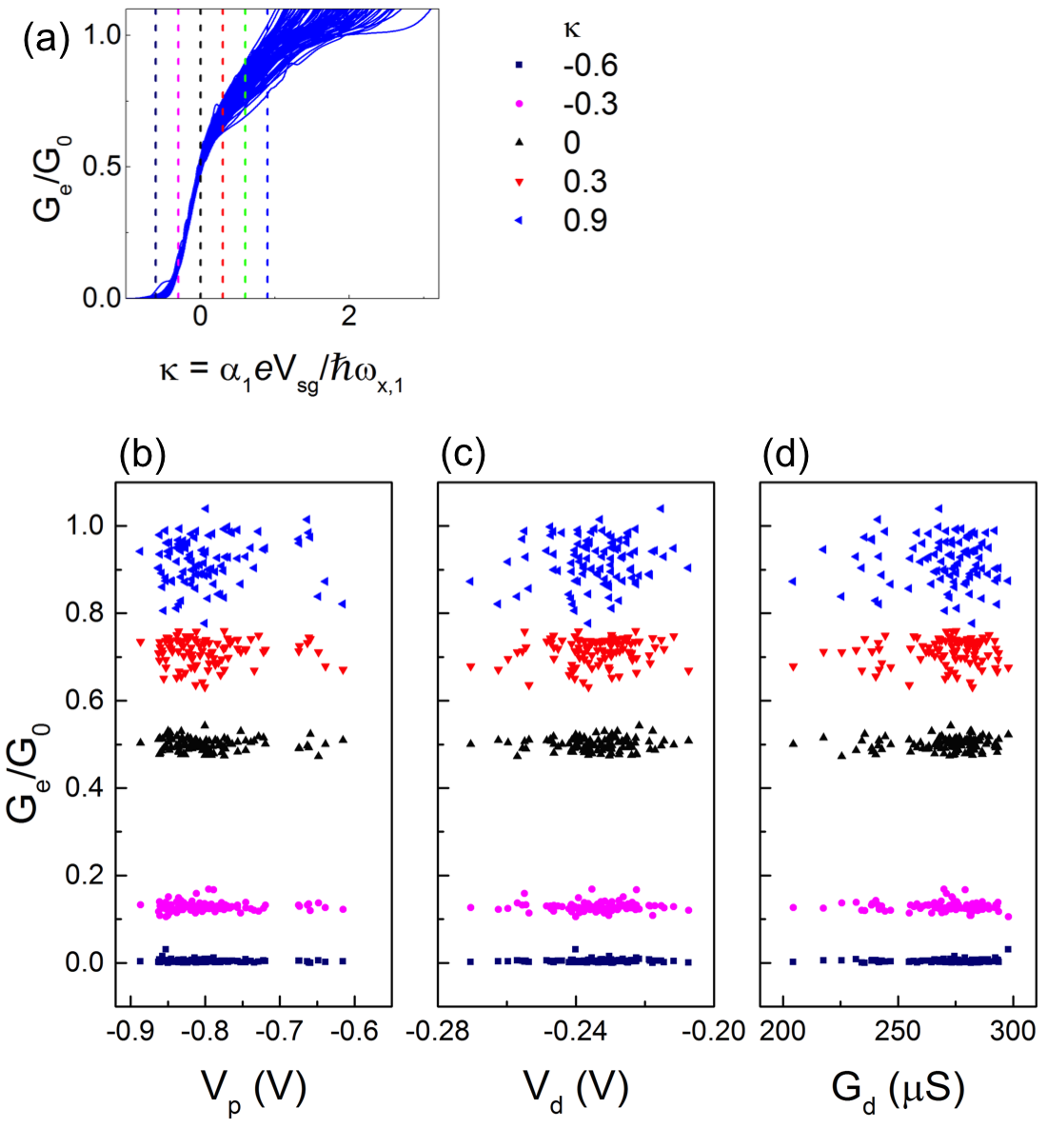}
\caption{\label{FigS1} (a) Full set of experimental conductance data $G_e/G_0$ from 98 split gates, as a function of $\kappa=\alpha_1 e V_{sg}/\hbar\omega_{x,1}$. Traces are offset horizontally to align $G_e/G_0=0.5$ with $V_{sg}=0$. The six vertical dashed lines show $\kappa=-0.6$ to $0.9$, in steps of $0.3$.
(b)--(d) Conductance $G_e/G_0$ as a function of pinch off voltage $V_p$, 1D definition voltage $V_d$, and definition conductance $G_d$, respectively. In each panel, data are shown for five fixed values of $\kappa$. From bottom-to-top, these correspond to vertical dashed lines (left-to-right) in panel (a) ($\kappa = 0.6$ is omitted for clarity). } 
\end{figure}

Figure S\ref{FigS1}(a) shows the experimentally measured conductance $G_e/G_0$ for 98 split gates, where $V_{sg}$ is scaled by $\alpha_1 e/\hbar\omega_{x,1}$ as described in the main article (traces are offset horizontally to align $G_e/G_0=0.5$ with $V_{sg}=0$). 
All traces collapse onto a similar curve below $G_e/G_0=0.5$, while variations exist for $0.5 <G_e/G_0<1$, due to the 0.7 anomaly. The six vertical dashed lines show $\kappa=-0.6$ to $0.9$, in steps of $0.3$.

We compare $G_e/G_0$ at fixed $\kappa$ against various properties of the 1D conductance trace.
Figure S\ref{FigS1}(b) shows $G_e/G_0$ as a function of pinch-off voltage $V_p$, at five values of $\kappa$ ($V_p$ refers to when $G_e$ has dropped to zero).
From bottom-to-top, these correspond to vertical dashed lines (left-to-right) in Fig. S\ref{FigS1}(a) ($\kappa=0.6$ data are omitted for clarity). 
Figures S\ref{FigS1}(c) and S\ref{FigS1}(d) show corresponding data as a function of the voltage ($V_d$) and conductance ($G_d$) at which a 1D channel is first defined, respectively.
There are no correlations between $G_e/G_0$ and any of these properties. 

In summary, the only clear correlation we have observed is between $G_e/G_0$ and $\hbar\omega_{x,1}$ [Fig. A2(a)]. This indicates that the barrier curvature is the most significant parameter in governing the behaviour of the 0.7 structure.\\

\begin{figure*}
\includegraphics[width=15cm,height=11cm,keepaspectratio]{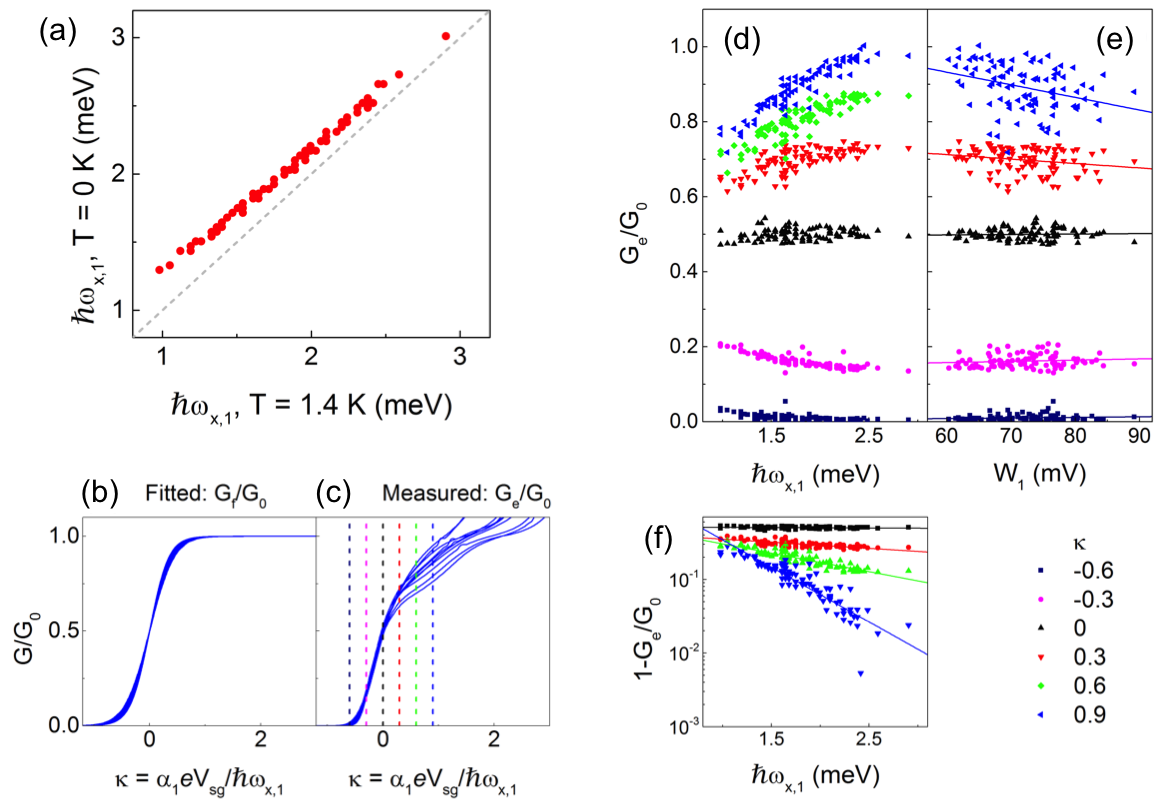}
\caption{\label{FigS2} 
(a) Comparison of $\hbar\omega_{x,1}$ values obtained 
using $T=0$ (vertical axis) or $T=1.4$ K (horizontal axis) in the fitting formula used to estimate $\hbar\omega_{x,1}$. For $T=0$ K, the estimate is systematically higher due to the absence of thermal broadening. The dashed line indicates the one-to-one correspondence between $x$ and $y$ axes, as a guide-to-eye.
(b) Fitted conductance $G_f/G_0$ at as a function of scaled gate voltage $\kappa$, for 98 split gates (data are first offset to align $G_f/G_0=0.5$ with $V_{sg}=0$). 
These data are calculated using $T=1.4$ K, and there is a small spread in the collapsed traces not present for the $T=0$ K case.
(c) Corresponding experimentally-measured conductance $G_e/G_0$ (for clarity, only ten traces are shown). The data are offset and scaled using the same parameters as $G_f$. 
Six vertical dashed lines show $\kappa$ from $-0.6$ to $0.9$ in steps of $0.3$.
(d) $G_e/G_0$ as a function of $\hbar\omega_{x,1}$, at fixed values of $\kappa$. From bottom-to-top, these correspond to vertical dashed lines (left-to-right) in panel (c). 
A small downward trend exists for data below $0.5G_0$, which arises from the small spread in the collapsed conductance traces. 
(e) Linear-log plot of $1-G_e/G_0$ as a function of $\hbar\omega_{x,1}$, for $\kappa = 0$ and above. The solid lines show linear fits to the data.
(f) Conductance $G_e/G_0$ as a function of width of the first conductance plateau $W_1$, where $W_1=\Delta V_{sg}$ between $G=0.5$ and $1.5G_0$. Data are shown for five fixed values of $\kappa$ ($\kappa = 0.6$ is omitted for clarity).
Linear fits are plotted for each $\kappa$. 
} 
\end{figure*}

\subsection{Performing the fitting with $T=1.4$ K for the first 1D subband}

For data presented in the main article, the fitted conductance for the first subband is calculated using $T=0$ K in the fitting formula. This is because experimental data show that the transition in $G$ from zero to $0.5G_0$ is temperature independent up to intermediate temperatures. Devices similar to those measured here show very little change up to $T=1.5$ K (see Fig. 4 of Ref.~\cite{Thomas1996S} for example). In the discussion below, we show that repeating our analysis using $T=1.4$ K (the $T$ at which the measurements were performed), gives similar results.

Figure S\ref{FigS2}(a) shows $\hbar\omega_{x,1}$ for $T=0$ K against $\hbar\omega_{x,1}$ for $T=1.4$ K. The dashed line shows the one-to-one correspondence of axes.
The values of $\hbar\omega_{x,1}$ are very similar; for $T=0$ K, $\hbar\omega_{x,1}$ is systematically slightly higher by about 0.2 meV (an increase is expected because there is no temperature broadening of the conductance trace).

Figures S\ref{FigS2}(b)--(f) reproduce Figs. A1(d), A1(e), and the whole of Fig. A2 from the main article. However, for the data show here, $T=1.4$ K is used in the fitting formula (for $n=1$) instead of $T=0$ K. Both sets of data are similar and show the same trends.

Figure S\ref{FigS2}(b) shows $G_f$ as a function of $\kappa$ for $T=1.4$ K.
The data collapse onto a curve similar to Fig. A1(d) from the main article. However, there are small variations which result in a spread near zero and $G_0$.
The corresponding experimental data $G_e$ for a subset of 10 devices are shown in Fig. S\ref{FigS2}(c).

Panels (d), (e) and (f) of Fig. S\ref{FigS2} show the dependence of $G_e/G_0$ on $\hbar\omega_{x,1}$ and $W_1$, when $T=1.4$ K is used for the fitting function.
The data are very similar for in both cases ($T=0$ and $1.4$ K). The only difference is that below $0.5G_0$, the data in Fig. S\ref{FigS2}(d) shows a slight upward trend (with an overall change of $\Delta G_e/G_0=0.07$ for $\kappa=-0.3$), as $\hbar\omega_{x,1}$ decreases from 3 to 1 meV.
This trend--not present in Fig. A2(a)--reflects the spread in $G_e$ visible in Fig. S\ref{FigS2}(c) below $0.5G_0$. 
There will be a corresponding small downward trend in Fig. S\ref{FigS2}(d) above $0.5G_0$. 

This illustrates a benefit of using $T=0$: The fitted conductance traces collapse onto an single curve [Fig. A2(a)], such that variations in transition widths as $G$ rises from zero to $G_0$ are completely removed. Since $\hbar\omega_{x,1}$ governs the width in gate voltage of this transition, the geometry dependence is effectively removed, and data from each split gate can be directly compared.


\begin{thebibliography}{MM}
\bibitem{NoteOmega} Parameter $\tilde{\Omega}_x$ is proportional to $\Omega_x$. Trends as a function of $\tilde{\Omega}_x$ can equally be described with reference to $\Omega_x$.
\end{thebibliography}
\end{document}